# Goal-oriented Setup and Usage of Custom-tailored Software Cockpits

Jens Heidrich, Jürgen Münch

Fraunhofer IESE, Fraunhofer Platz 1, 67663 Kaiserslautern, Germany
{jens.heidrich, juergen.muench}@iese.fraunhofer.de

**Abstract.** Software Cockpits, also known as Software Project Control Centers, support the management and controlling of software and system development projects and provide means for quantitative measurement-based project control. Currently, many companies are developing simple control dashboards that are mainly based on Spreadsheet applications. Alternatively, they use solutions providing a fixed set of project control functionality that cannot be sufficiently customized to their specific needs and goals. Specula is a systematic approach for defining reusable, customizable control components and instantiate them according to different organizational goals and characteristics based on the Quality Improvement Paradigm (QIP) and GQM. This article gives an overview of the Specula approach, including the basic conceptual model, goal-oriented measurement, and the composition of control components based on explicitly stated measurement goals. Related approaches are discussed and the use of Specula as part of industrial case studies is described.

## 1 Introduction

The complexity of software development projects continues to increase. One major reason is the ever-increasing complexity of functional as well as non-functional software requirements (e.g., reliability or time constraints for safety-critical systems). The more complex the requirements, the more people are usually involved in meeting them, which further increases the complexity of controlling and coordinating the project. This, in turn, makes it even harder to develop the system according to plan (i.e., matching time and budget constraints). Project control issues are very hard to handle. Many software development organizations still lack support for obtaining intellectual control over their software development projects and for determining the performance of their processes and the quality of the produced products. Systematic support for detecting and reacting to critical project states in order to achieve planned goals is often missing [15].

Companies have started to introduce so-called software cockpits, also known as Software Project Control Centers (SPCC) [15] or Project Management Offices (PMO) [16], for systematic quality assurance and management support. A software cockpit is comparable to an aircraft cockpit, which centrally integrates all relevant information for monitoring and controlling purposes. A project manager can use it to get an overview of the project state and a quality assurance manager can use it to check the quality of the software product. In addition to these primary users of an SPCC, basically any role of a project may profit from making direct or indirect use of the SPCC functionality. For instance, a developer can use the SPCC to keep track of code quality or to trace quality issues. The benefit provided by an SPCC for a certain project role depends on the functionality and services offered. However, the needs with respect to project control differ between different organizations, projects, and roles. They depend on organizational goals (business goals), process maturity, the experience of the project team, and many other factors. For instance, for multi-disciplinary, distributed software development, measurement data has to be collected from different sources (locations) and formats. In this case, integration of data is crucial for getting a consistent picture of the project state.

In general, an important success factor in the software engineering domain is that these solutions are customized to the specific goals, organizational characteristics and needs, as well as the concrete project environment. Specula (*lat.* watch tower) is an approach for composing project control functionality out of reusable control components [7], [8]. It was mainly developed at the Fraunhofer Institute for Experimental Software Engineering (IESE) and makes use of the Quality Improvement Paradigm (QIP) for integrating project control activities into a continuous improvement cycle. Furthermore, the GQM approach [2] is used for explicitly specifying measurement goals for project control.

Section 2 of the article presents related work in the field of software project control centers and key performance indicators for project control. Section 3 introduces the Specula approach, describes the underlying conceptual model and its relationship to goal-oriented measurement, and finally presents the basic steps of the methodology for composing control components (encapsulated, packaged techniques for project control) based on explicitly defined measurement goals. Section 4 presents first empirical evaluation results based on industrial case studies conducted. The article concludes with a brief summary and discussion of future work.

## 2    Related Work

An overview of the state of the art in Software Project Control Centers can be found in [15]. The scope was defined as generic approaches for online data interpretation and visualization on the basis of past experience. However, project dashboards were not included in this overview. In practice, many companies develop their own dashboards (mainly based on Spreadsheet applications) or use dashboard solutions that provide a fixed set of predefined functions for project control (e.g., deal with product quality only or solely focus on project costs) and are very specific to the company for which they were developed. Most of the existing, rather generic, approaches for control centers offer only partial solutions. Especially purpose- and role-oriented usages based on a flexible set of techniques and methods are not comprehensively supported. For instance, SME (Software Management Environment) [10] offers a number of role-oriented views on analyzed data, but has a fixed, built-in set of con-trol indicators and corresponding visualizations. The SME successor WebME (Web Measurement Environment) [19] has a scripting language for customizing the interpretation and visualization process, but does not provide a generic set of applicable controlling functions. Unlike Provence [13] and PAMPA [18], approaches like Amadeus [17] and Ginger2 [20] offer a set of purpose-oriented controlling functions with a certain flexibility, but lack a role-oriented approach to data interpretation and visualization.

The indicators used to control a development project depend on the project's goals and the organizational environment. There is no default set of indicators that is always used in all development projects in the same manner. According to [14], a "good" indicator has to (a) support analysis of the intended information need, (b) support the type of analysis needed, (c) provide the appropriate level of detail, (d) indicate a possible management action, and (e) provide timely information for making decisions and taking action. The concrete indicators that are chosen should be derived in a systematic way from the project goals [12], making use of, for instance, the Goal Question Metric (GQM) approach. Some examples from indicators used in practice can be found in [1]. With respect to controlling project cost, the Earned Value approach provides a set of commonly used indicators and interpretation rules. With respect to product quality, there exists even an ISO standard [11]. However, the concrete usage of the proposed measures depends upon the individual organization. Moreover, there is no unique classification for project control indicators. One quite popular classification of general project management areas is given by the Project Management Body of Knowledge (PMBoK) [16]. The PMBoK distinguishes between nine areas, including project time, cost, and quality management.

The ideas behind GQM and the Quality Improvement Paradigm (QIP) [2] are well-proven concepts that are widely applied in practice today. An approach based on GQM and QIP to create and maintain enhanced measurement plans, addressing data interpretatation and visualization informally, is presented in [5]. Moreover, related work in this field is presented.

## 3   The Specula Approach

Specula is a state-of-the-art approach for project control. It interprets and visualizes collected measurement data in a goal-oriented way in order to effectively detect plan deviations. The control functionality provided by Specula depends on the underlying goals with respect to project control. If these goals are explicitly defined, the corresponding functionality is composed out of packaged, freely configurable control components. Specula provides four basic components: (1) a logical architecture for implementing software cockpits [15], (2) a conceptual model formally describing the interfaces between data collection, data interpretation, and data visualization [9], (3) an implementation of the conceptual model, including a construction kit of control components [4], and (4) a methodology of how to select control components according to explicitly stated goals and customize the SPCC functionality [9].

The methodology is based on the Quality Improvement Paradigm (QIP) and makes use of the GQM approach [2] for specifying measurement goals. QIP is used to im-plement a project control feedback cycle and make use of experiences and knowledge gathered in order to reuse and customize control components. GQM is used to drive the selection process of finding the right control components according to defined goals. Large parts of the approach are supported by a corresponding prototype tool, called Specula Project Support Environment (PSE), which is currently also being used as part of industrial case studies (see Section 4 and [4]). Specula basically addresses the following roles that make use of the provided functionality:

- *Primary Users:* Project manager, quality assurance manager, and controller who mainly use an SPCC to control different aspects of the software development project and initiate countermeasures in case of deviations and risks.

- *Secondary Users:* Developers and technical staff who use an SPCC to enter measurement data as well as to detect root causes for deviations and risks.

- *Administrators:* Administrators who have to install and maintain an SPCC.

- *Measurement Experts:* Experts who define measurement goals, support derivation of control components, and help to customize and effectively use the SPCC.

Section 3.1 gives a brief overview of the conceptual model upon which Specula is built. Section 3.2 addresses the connection of the conceptual model to goal-oriented measurement, and Section 3.3 provides a brief overview of all steps necessary to apply the Specula approach as a whole.

### 3.1     Cockpit Concepts

The conceptual model of the Specula approach formalizes the process of collecting, interpreting, and visualizing measurement data for software project control. The derived structure for operationally controlling a development project is called a *Visualization Catena (VC)* [7], which defines components for automatically and manually collecting measurement data, processing and interpreting these data, and finally visualizing the processed and interpreted data. The processing and interpretation of collected measurement data is usually related to a special measurement purpose, like analyzing effort deviations, or guiding a project manager. A set of techniques and methods (from the repository of control components) is used by the VC for covering the specified measurement purpose. The visualization and presentation of the processed and collected measurement data is related to roles of the project that profit from using the data. The VC creates a set of custom-made controlling views, which presents the data according to the interests of the specified role, such as a high-level controlling view for a project manager, and a detailed view of found defects for a quality assurance manager. The whole visualization catena has to be adapted in accordance with the context characteristics and organizational environment of the software development project currently being controlled.

Fig. 1 gives an overview of all VC components and their corresponding types. Specula distinguishes between the following five components on the type level from which a concrete VC is instantiated for a certain project:

(T1) Data types describe the structure of incoming data and data that is further processed by the VC. For instance, a time series (a sequence of time stamp and corresponding value pairs) or a project plan (a hierarchical set of activities having a start and end date and an effort baseline) could be logical data types that could either be directly read-in by the system or be the output of a data processing function.

(T2) Data access object packages describe the different ways concrete data types may be accessed. For instance, an XML package contains data access objects for reading data (having a certain data type) from an XML file, writing data to an XML file, or changing the contents of an XML file. A special package may be used, for instance, to automatically connect to an effort tracking system or bug tracking data base. A data access object contains data type-specific parameters in order to access the data repositories.

(T3) Web forms describe a concrete way of managing measurement data manually, involving user interaction. A web form manages a concrete data type. For instance, new data may be added, existing data may be changed or completely removed. A web form also refers to other data types that are needed as input. For instance, in order to enter effort data manually, one needs the concrete activities of the project for which the effort is tracked. Web forms are needed if the data cannot be automatically retrieved from an external data source.

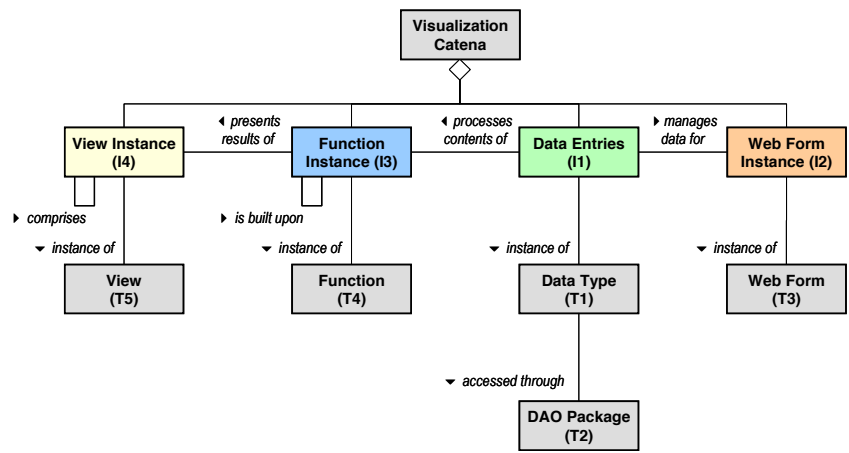

**Fig. 1.** Overview of the elements of the conceptual model. A view instance presents the results of a data processing function, which in turn processes the contents of data entries for which data is provided by a web form instance.

(T4) Functions represent a packaged control technique or method, which is used to process incoming data (like Earned Value Analysis, Milestone Trend Analysis, or Tolerance Range Checking). A function needs different data types as input, produces data of certain data types as output, and may be adapted to a concrete context through a set of parameters.

(T5) Views represent a certain way of presenting data, like drawing a two-dimensional diagram or just a table with a certain number of rows and columns. A view visualizes different data types and may refer to other views in order to create a hierarchy of views. The latter may, for instance, be used to create a view for a certain project role consisting of a set of sub-views.

In addition, the following components are distinguished on the instances level:

(I1) Data entries instantiate data types and represent the concrete content of measurement data that are processed by the SPCC. We basically distinguish between external and internal data. External data must be read-in or imported from an external location, or manually entered into the system. Each external data object has to be specified explicitly by a data entry containing, for instance, the start and end time and the interval at which the data should be collected. In addition, the data access object package that should be used to access the external data has to be specified. Internal data are the outcome of functions. They are implicitly specified by the function producing the corresponding data type as output and therefore need no explicit specification and representation as data entry. External as well as internal data may be used as input for instances of functions or views if their corresponding data types are compatible.

(I2) Web form instances provide web-based forms for manually managing measurement data for data entries. All mandatory input data type slots of the instantiated web form have to be filled with concrete data entries and all mandatory parameters have to be set accordingly.

(I3) Function instances apply the instantiated function to a certain set of data entries filling the mandatory input slots of the function. A function instance processes (external and internal) data and produces output data, which could be further processed by other function instances or visualized by view instances. All mandatory function parameters have to be set accordingly.

(I4) Finally, view instances apply the instantiated view to a certain set of data entries filling the corresponding mandatory data type slots of the view. A view instance may refer to other view instances in order to build up a hierarchy of views.

Each component of a VC and its corresponding type contains explicitly specified checks that may be used to test whether the specification is complete and consistent, whether data are read-in correctly, whether function instances can be computed accurately, and whether view instances can be created successfully. A visualization catena consists of a set of data entries, each having exactly one active data access object for accessing incoming data, a set of web form instances for managing the defined data entries, a set of function instances for processing externally collected and internally processed data, and finally, a set of view instances for visualizing the processing results. A formal specification of all components may be found in [6].

### 3.2   Mapping Cockpit Concepts to GQM

For a goal-oriented selection of control components, a structured approach is needed that describes how to systematically derive control components from project goals and characteristics. GQM provides a template for defining measurement goals, systematically derives questions that help to make statements about the goals, and finally derives metrics in order to help answer the stated questions. In order to complete such a measurement plan for a concrete project, each metric can be further described by a data collection specification (DCS) basically making statements on who or which tool has to collect the measurement data at which point in time of the project from which data source. In [8], usage scenarios on how to derive a GQM plan from a control goal and how to define a VC that is consistent with the defined goals are described.

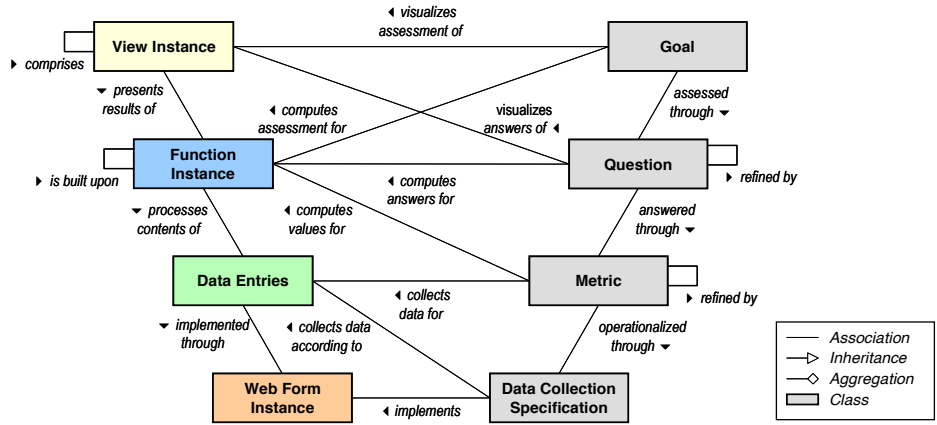

**Fig. 2.** Mapping the conceptual model to the GQM paradigm. On the left side, one can see the components of the visualization catena. On the right side, one can see the structure of a GQM model and a corresponding data collection specification.

Fig. 2 presents an overview of all relationships between a GQM plan, its DCS, and a visualization catena (cf. [9]):

- Data entries collect measurement data for GQM metrics according to the DCS. If the data has to be collected manually, a web form instance is used to implement the DCS in addition. For instance, if the DCS states that the start and end date of an activity shall be collected from an MS Project file, a corresponding data entry is defined and a web form instance implements importing the project plan from the file.

- Function instances compute metric values if a metric has to be computed from other metrics. For instance, if a cost performance index is computed for an Earned Value Analysis, the budgeted costs of work performed and the actual costs of work performed are needed. A function instance could also compute answers for GQM questions by taking into account all metrics assigned to the question and applying an interpretation model to all metric values. In analogy, a function instance could assess the attainment of a GQM goal by assessing the answers of all assigned questions using an interpretation model.

- View instances visualize the answers to GQM questions. A chart is produced or tables are displayed illustrating the metric results of the corresponding questions and the interpretation model used to answer the question. For instance, the cost performance and schedule performance index could be visualized as a line chart in which good and bad index values are marked accordingly. A view instance could also visualize the assessment of the GQM goal.

### 3.3 Composing Control Components

Specula is largely based on the Quality Improvement Paradigm (QIP). The basic phases and steps are as follows:

*Phase I: Characterize Control Environment:* First, project stakeholders characterize the environment in which project control shall be applied in order to set up a corresponding measurement program that is able to provide a basis for satisfying all needs.

- *Describe the project context.* Important characteristics for setting up project control mechanisms have to be defined.

- *Discuss the overall organization.* Organizational characteristics have to be clarified. This includes roles and responsibilities, potential stakeholders, like managers of the organization, project managers, quality assurance manager, developers, and team organization.

*Phase II: Set Control Goals:* Then, measurement goals for project control are defined and metrics are derived determining what kind of data to collect.

- *Elicit control goals.* The Specula approach makes use of GQM in order to define measurement goals in a structured way. GQM already provides a systematic approach for defining measurement goals, systematically derives questions that help to make statements about the goals, and finally derives metrics in order to help answer the stated questions.

- *Clarify relations to higher-level goals.* The relation to higher-level goals should be modeled. For this purpose, all measurement goals are connected to higher-level software and business goals using the GQM$^+$Strategies® approach [3].

- *Derive indicators.* Based on the measurement goals defined for project control, questions and metrics have to be derived using GQM.

- *Define GQM model.* A GQM model is created containing the project-specific measurement goals, corresponding questions that make statements about achieving goals, and metrics that support answering the questions.

*Phase III: Goal-oriented Composition:* Next, a visualization catena is composed based on the defined goals in order to provide online feedback on the basis of the data collected during project execution. More details about this process can be found in [9].

- *Derive measurement plan.* A comprehensive measurement plan has to be derived based on the GQM model, including a data collection specification.

- *Define interpretation models.* Interpretation models are used to basically aggregate measurement data in order to answer a GQM question or make a statement about achieving a GQM goal.

- *Derive data entries and web form instances.* Next, matching data types are identified based on the metric definition, the object to be measured and the quality attribute. For each simple metric (which is not computed from other metrics), instantiate the data type and create a corresponding data entry. The data collection specification is used to determine the start time, end time, and interval when the data should be collected. If the metric has to be collected manually, a web form is identified based on the data source and the instantiated web form is attached to the data entry.

- *Derive function instances for complex metrics*. For each complex metric (which is computed from other metrics), a function is identified that is able to compute the metric based on the metric definition, the object to be measured, and the quality attribute. The identified functions are instantiated by first filling all input data slots with data entries or results of other function instances. Then, the function instances are parameterized according to the metric definition.
- *Derive function instances for GQM questions*. If an interpretation model is described in the GQM plan that defines how to formally answer a question, a function implementing this model is identified based on the object and quality attribute addressed in order to compute the answers to the question. The functions are instantiated by filling all input data slots with data entries or results of other function instances assigned to the question. The function instances are parameterized according to the interpretation model.
- *Derive view instances for GQM questions*. The answers to the question are visualized by identifying a set of views based on the kind of answers to the question and the data visualization specifications of the measurement plan (if any). The identified views are instantiated by filling all input data slots with data entries or results of function instances assigned to the question. The view instances are parameterized according to the data presented (e.g., title and axis description, size, and color).
- *Derive function instances for GQM goals*. If an interpretation model is described in the GQM plan that defines how to formally assess goal attainment, a function implementing this model is identified and instantiated based on the object and quality focus addressed in order to attain the measurement goal.
- *Derive view instances for GQM goals*. Goal attainment is visualized by identifying and instantiating a set of views based on the kind of assessment of the goal and the data visualization specifications of the measurement plan (if any).
- *Check consistency and completeness*. After defining the whole visualization catena for controlling the project, the consistency and completeness of the mapping process are checked.
- *Configure SPCC*. If the visualization catena is defined and checked, it has to be transferred to a corresponding tool (Specula tool prototype).
- *Provide training*. Training is provided for all SPCC users in order to guarantee the effective usage of the SPCC.

*Phase IV: Execute Project Control Mechanisms:* Once the visualization catena is specified, a set of role-oriented views are generated by the SPCC for controlling the project based on the specified visualization catena. If a plan deviation or project risk is detected, its root cause must be determined and the control mechanisms have to be adapted accordingly.

- *Perform data collection*. The SPCC users have to perform data collection activities according to the measurement plan defined.
- *Use control views for GQM questions*. The SPCC users have to use the view instances offered to get answers for the GQM questions of their GQM model.

- *Use control views for GQM goals*. The SPCC users have to use the view instances offered to get a general answer with respect to achieving a certain goal of the GQM models.
- *Check SPCC functionality*. The SPCC users should check the correct functionality of the Project Control Center regularly.

*Phase V: Analyze Results:* After project completion, the resulting visualization catena has to be analyzed with respect to plan deviations and project risks detected intime, too late, or not detected at all. The causes for plan deviations and risks that have been detected too late or not all have to be determined.

- *Analyze plan deviations and project risks*. The complete lists of plan deviations and project risks have to be analyzed after the end of the project.
- *Analyze measurement plan*. For all deviations and risks that were not detected at all, the measurement plan has to be analyzed with respect to missing goals or other missing parts of the GQM models.
- *Analyze interpretation models*. For all deviations and risks that were not detected at all or that were detected too late, the interpretation models have to be checked to see whether they work as intended or whether metrics or answers to questions need to be interpreted in a different way.
- *Analyze visualization catena*. For all deviations and risks that were detected too late, the components of the visualization catena that helped in detecting them have to be analyzed to see whether they can be improved to support earlier detection in future projects.

*Phase VI: Package Results:* The analysis results of the visualization catena that was applied may be used as a basis for defining and improving control activities for future projects (e.g., selecting the right control techniques and data visualizations, choosing the right parameters for controlling the project).

## 4     Empirical Evaluation and Usage Example

The evaluation of the Specula approach is currently being conducted in the context of several industrial case studies as part of the Soft-Pit research project funded by the German Federal Ministry of Education and Research (http://www.soft-pit.de). The project focuses on getting experience and methodological support for operationally introducing control centers into companies and projects. The project includes performing several industrial case studies with German companies from different domains, in which the developed control center and its deployment are evaluated. The project is mainly organized intro three iterations focusing on different controlling aspects. An application of Specula in the first iteration showed the principal applicability of the VC concept in an industrial environment. Results can be found in [4]. The second iteration focused on three aspects: (a) perceived usefulness and ease of use of the approach, (b) found plan deviations and project risks, and (c) costs for setting up and applying an SPCC. Those aspects were evaluated in four industrial case studies, in which the Specula prototype tool was used to control the software development project. The system was perceived as useful and easy to use. However, the degree of usefulness depended on the group of users: the benefits for secondary users were limited. Usefulness also varied across different organizations; this may be related to the different control mechanisms used before introducing an SPCC. Preliminary results show that following a structured process for setting up an SPCC also does result in a significantly improved detection rate of plan deviations and project risks. The costs for setting up and applying an SPCC were around 10% of the overall development effort for a medium-sized project (10 team members). In the following, the basic steps of the method are illustrated using data from a practical course conducted at the University of Kaiserslautern in which the Specula project control approach was applied.

*Phase I: Characterize Control Environment:* The aim was to develop mobile services for creating a virtual office of the future. There were 17 team members. The project manager and quality assurance manager should use an SPCC to control different aspects of the project. In addition, an administrator (not part of the project team) was provided who was familiar with the SPCC tool.

*Phase II: Set Control Goals:* A measurement expert conducted structured interviews with the project manager and quality assurance manager in order to retrieve the measurement goals with respect to project control that are to be achieved:

- Analyze the project plan for the purpose of monitoring the consistency of the plan from the point of view of the project manager.
- Analyze the project plan for the purpose of comparing the actual effort with the planned effort from the point of view of the project manager.
- Analyze the project plan for the purpose of monitoring schedule adherence from the point of view of the project manager.
- Analyze the project plan for the purpose of monitoring effort tracking regularity from the point of view of the project manager.
- Analyze the source code for the purpose of monitoring the quality from the point of view of the quality assurance manager.
- Analyze the defect detection activities for the purpose of monitoring their efficiency from the point of view of the quality assurance manager.

*Phase III: Goal-oriented Composition:* A visualization catena was created for the GQM goals above. For example, if the goal is to evaluate the effort plan with respect to plan deviation, the corresponding control components can be selected as follows. Fig. 3 presents the GQM model for this goal on the left side and the corresponding excerpt of the resulting VC on the right side. The one and only question asked was about absolute effort deviation per activity. A complex metric defined the deviation as the amount that an actual effort value is above an effort baseline. Three simple metrics were consequently defined and operationalized by corresponding data collection specifications. The baseline should be extracted from a project plan stored in an MS project file, so a corresponding web form collecting project plan information and data types representing the project activities and the effort baseline were instantiated. The actual effort data should be extracted from the company-wide effort tracking system including effort per person and activity. A data type was instantiated that accesses the tracking system using a corresponding data access object. A function was applied to aggregate the effort data for each activity across all persons. In order to compute the complex metric "effort plan deviation", a tolerance range checking function was applied that computes the deviation accordingly. Finally, a view was instantiated in order to graphically display the results of the assigned function instances and data entries. Fig. 4 presents the complete visualization catena that was derived for all goals defined as outputted by the Specula prototype tool (instantiation of the concepts shown in Fig. 1). As can be seen, the logical dependency of components is quite high, even for a limited number of control components. The excerpts of the VC discussed above are highlighted accordingly.

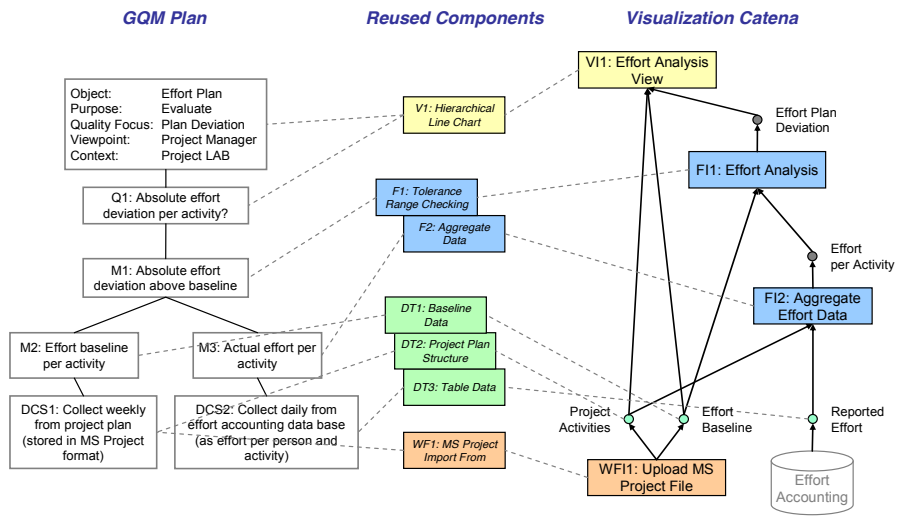

**Fig. 3.** Composing the VC from reusable components. The left side shows the GQM plan to be implemented by an SPCC. According to the information specified in the GQM plan, components are identified from a reuse repository and instantiated in order to create a visualization catena.

*Phase IV: Execute Project Control Mechanisms:* Fig. 5 presents a visualization of the effort controlling view generated by the Specula prototype tool. During the execution of the project, the team members entered their effort data using the corresponding Specula web form. The project manager regularly updated the project plan using MS Project and imported the plan into the SPCC. The quality assurance manager used a static code analysis tool to analyze code quality and imported a corresponding report into the SPCC.

*Phase V: Analyze Results:* General deviations from the effort baseline were detected including, but not limited to, that the requirements phase took a lot more effort than planned. The project manager updated the project plan accordingly. In addition, if we assume that a negative milestone trend was not detected at all, an important milestone might have been missed.

*Phase VI: Package Results:* If we assume that the control component for detecting milestone trends used a wrong parameter setting, it will have to be adapted for future use in subsequent projects.

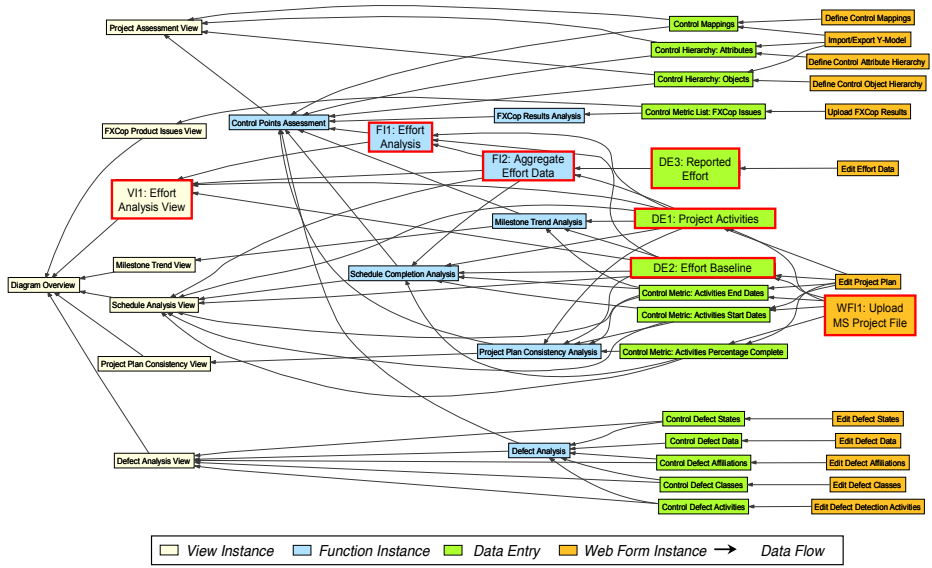

**Fig. 4.** Example visualization catena. One can see all input and output data of all control components used for constructing the VC. 13 web form instances provide input for 15 data entries, which are processed by 8 function instances, and visualized by 8 view instances.

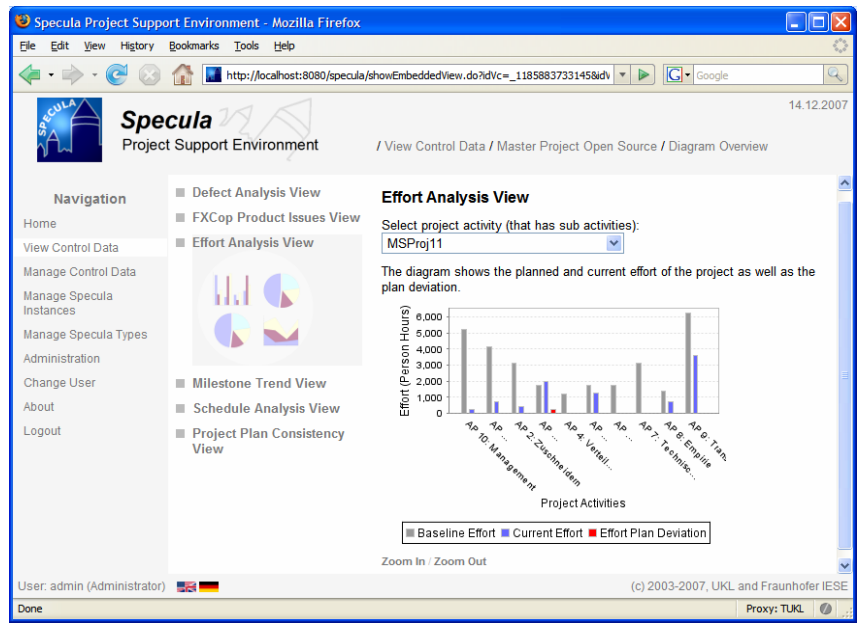

**Fig. 5.** User interface of the Specula prototype tool. On the left side, one can see the overall navigation bar. The menu close to the navigation bar displays all available views for controlling the project. On the right side, one can see the selected view for analyzing effort data.

## 5     Conclusion and Future Work

The article presented the Specula controlling approach for setting up a project control mechanism in a systematic and goal-oriented way, profiting from experiences gathered. Reusable control components were defined and instantiated to illustrate how to define measurement-based project control mechanisms and instantiate them for the software development projects of a concrete organization. A high-level process was shown that provided guidance on how to select the right control components for data collection, interpretation, and visualization based on explicitly defined measurement goals. Moreover, a simple example was presented of how to apply generically defined control components. The Specula approach implements a dynamic approach for project control; that is, measures and indicators are not predetermined and fixed for all projects. They are dynamically derived from measurement goals at the beginning of a development project. Existing control components can be systematically reused across projects or defined newly from scratch. Data is provided in a purpose- and role-oriented way; that is, a certain role sees only measurement data visualizations that are needed to fulfill the specific purpose. Moreover, all project control activities are defined explicitly, are built upon reusable components, and are systematically performed throughout the whole project. A context-specific construction kit is provided, so that elements with a matching interface may be combined. The qualitative benefits of the approach include: being able to identify and reduce risks related to introducing software cockpits, being more efficient in setting up and adapting project controlling mechanisms, allowing for more transparent decision-making regarding project control, reducing the overhead of data collection, increasing data quality, and, finally, achieving projects that are easier to plan and to control.

Further development and evaluation of the approach will take place in the context of the Soft-Pit project. Future work will also concentrate on setting up a holistic control center that integrates more aspects of engineering-style software development (e.g., monitoring of process-product dependencies and linking results to higher-level goals). The starting point for setting up such a control center are usually high-level business goals, from which measurement programs and controlling instruments can be derived systematically. Thus, it would be possible to transparently monitor, assess, and optimize the effects of business strategies performed.

## Acknowledgements

This work was supported in part by the German Federal Ministry of Education and Research (Soft-Pit Project, No. 01ISE07A). We would also like to thank Sonnhild Namingha from Fraunhofer IESE for reviewing a first version of this article.